\documentclass[pra,amsmath,amssymb,notitlepage, twocolumn,superscriptaddress,longbibliography,nofootinbib,floatfix]{revtex4-2}

\usepackage{geometry}
\usepackage{tensor}
\usepackage{amsmath}
\usepackage{amssymb, mathtools}

\usepackage{enumerate}
\usepackage{hyperref}
\usepackage{lipsum}
\usepackage{slashed}
\usepackage{braket}
\usepackage{yfonts}
\usepackage{graphicx}
\usepackage{dcolumn}
\usepackage{bm}
\usepackage{footmisc}

\begin{document}

\preprint{APS/123-QED}

\title{Fixed points of classical gravity coupled with a Standard-Model-like theory}

\author{Latham Boyle}\affiliation{\Edinburgh}\affiliation{\PI}
\author{Neil Turok}\affiliation{\Edinburgh}\affiliation{\PI}
\author{Vatsalya Vaibhav}\affiliation{\Edinburgh}
\newcommand*{\Edinburgh}{Higgs Centre for Theoretical Physics, James Clerk Maxwell Building, Edinburgh EH9 3FD, UK} 
\newcommand*{\PI}{Perimeter Institute for Theoretical Physics, Waterloo, Ontario N2L 2Y5, Canada}

\begin{abstract}
Coupling quantum field theory (QFT) \!-\! even free QFT \!-\! to gravity leads to well-known problems. 
In particular, the stress tensor $T_{\mu\nu}$ (gravity's source) and 
its correlators typically diverge in the UV, creating a conflict between the wildly inhomogeneous spacetime we expect quantum mechanically and the weakly-curved, macroscopic spacetime we observe.
Are there QFTs for which these divergences cancel?  Here, for simplicity, we consider free quantum fields on a classical curved background.  The aforementioned divergences are related to the  running of the gravitational couplings. We calculate the corresponding beta functions, identifying a special class of QFTs with 
UV fixed points at which $\langle T_{\mu\nu}\rangle$ and all its 
correlators $\langle T\ldots T\rangle$ are UV finite.  An intriguing example is a theory like the Standard Model (including right-handed neutrinos) with $12$ gauge fields, $3$ generations of $16$ Weyl fermions and $36$ four-derivative (Fradkin-Tseytlin) scalars.  
In the infrared, this theory has a positive Newton's constant $G$ and an arbitrarily small cosmological constant $\Lambda$.

\end{abstract}

\maketitle 


\section{Introduction}  

Consider a quantum field theory (QFT) on a classical, curved spacetime background~\cite{DeWitt:1975ys, Birrell:1982ix, Mukhanov:2007zz}.
To renormalize it, local counterterms depending on the background metric and curvature are required in the Lagrangian. The coefficients of these ``gravitational" terms undergo renormalization group (RG) flow~\cite{Stueckelberg:1952hma, Gell-Mann:1954yli, Kadanoff:1966wm, Callan:1970yg, Symanzik:1970rt, Wilson:1973jj} (although we are not quantizing gravity) ~\cite{DeWitt:1975ys, Birrell:1982ix,  Mukhanov:2007zz, wald1994quantum, Hollands:2002ux, Percacci:2005wu}. Can this flow have fixed points where the corresponding gravitational beta functions vanish? If so, what is their significance?  

Recall that in flat spacetime, a QFT is only UV-complete if it flows to a UV fixed point. If it does not, it can only be a low-energy effective description. Even if it {\it is} UV-complete, problems arise in coupling it to gravity. Correlators of the stress tensor $T_{\mu\nu}$, gravity's source, are typically afflicted by UV divergences~\cite{DeWitt:1975ys,Osborn:1993cr}, raising a profound puzzle: why is spacetime apparently so gently curved when the source for gravity diverges so badly due to short wavelength fluctuations?  And does a UV/continuum limit for spacetime even exist?  One possibility is that QFT is only valid up to a cutoff of order the Planck scale $m_{Pl}$~\cite{Caron-Huot:2024lbf}. Even then, quantum zero-point fluctuations in the stress tensor would be expected to be $\sim m_{Pl}^4$, giving rise to a wildly curved spacetime at odds with the well-ordered, macroscopic universe we observe.  

In this Letter, we point out an alternative. We identify a special class of QFTs which possess gravitational fixed points (fixed points at which all beta functions, {\it including} the gravitational ones, vanish). As we shall explain (see  Sec.~\ref{Sec:Running}), the stress tensor correlators in these theories are completely free of UV divergences. Thus, {\it a priori}, these QFTs have a better chance of coupling sensibly to gravity.
Furthermore, one of them is intriguingly
close to the Standard Model (SM) of particle physics. 

Previous work \cite{Percacci:2005wu} computed the gravitational $\beta$ functions due to ordinary quantum matter fields, {\it i.e.}, gauge fields, spinor fields, and 2-derivative scalars. From these results, for the $\beta$ function of the $R^2$ coupling to vanish, the matter content must be restricted to {\it conformally coupled} fields.\footnote{Other indications that conformally coupled fields couple more consistently to gravity have previously been emphasized {\it e.g.}\ in Section~2.4 in \cite{DeWitt:1975ys}.}  But {\it conventional} conformally coupled fields are insufficient to cancel the  other gravitational $\beta$ functions.  

However, a new ingredient changes the story: Fradkin and Tseytlin \cite{Fradkin:1982xc, Fradkin:1981jc} (and later Paneitz \cite{paneitz2008quartic}) noticed that there are actually {\it two} ways to conformally couple a scalar field to gravity in four dimensions: via the conventional two-derivative (``Klein-Gordon" or ``KG") action \cite{Penrose:1964ge, Callan:1970ze}, or via a four-derivative (``Fradkin-Tseytlin" or ``FT") action \cite{Fradkin:1982xc, Fradkin:1981jc}.  Here, we extend the calculation of the gravitational beta functions to include FT scalars.  Gravitational fixed points then exist, but are rare.

In particular, let $n_{1}$, $n_{1/2}$, $n_{0}$ and $n_{0}'$ denote the number of gauge fields, Majorana or Weyl spinors, KG scalars and FT scalars, respectively.  As we shall see, canceling the gravitational beta functions requires $n_{1/2}=4n_{1}$, $n_{0}'=3n_{1}$, and $n_{0}=0$ (no fundamental KG scalars).  Since an FT scalar has twice as many degrees of freedom as a KG scalar \cite{Fradkin:1981iu}, such a theory has equal numbers of bosonic and fermionic degrees of freedom. Furthermore, the ratio of vector, spinor, and scalar degrees of freedom is $1:4:6$, as in maximal ($\mathcal{N}=4$) flat spacetime supersymmetry. 

One such fixed point is particularly intriguing from a phenomenological standpoint.  Consider the SM's $n_{1}=8+3+1=12$ gauge fields and $n_{1/2}=3\times16=48$ Weyl spinors (including right-handed neutrinos). As previously noted~\cite{Boyle:2021jaz}, adding $n_{0}'=36$ FT scalars (and no KG scalars, as appropriate if the SM Higgs is composite), cancels the leading-order vacuum energy and $a$ and $c$ Weyl anomalies. Furthermore, FT scalars can provide a non-inflationary explanation of the observed spectrum of primordial density perturbations \cite{Turok:2023amx}.  The renormalization group (RG) analysis here is more powerful, allowing us to study the flow of Newton's constant $G$.\footnote{One must carefully distinguish different definitions of the running of the gravitational couplings, which have different meaningins, and confounding them can lead to confusion.  In particular, in this paper we study the running defined by the Exact Renormalization Group Equation (ERGE) for the effective action \cite{wetterich1993exact} (see Sec.~\ref{beta_fn_sec} below), and with this definition the dimensionful gravitational couplings {\it do} run.  By contrast, using a different definition (the running of Lorentzian scattering amplitudes with respect to external momenta) it has been argued that the gravitational couplings do {\it not} run \cite{Donoghue:2024uay}.}
We find $G$ is constant in the IR, with the correct sign, provided the continuation to Euclidean signature is performed with due care, in accordance with realistic hot big bang cosmology.

\section{Scale-invariant matter on a classical spacetime background}  

Consider the Euclidean action $S_{{\rm mat}}$ for a collection of free fields: $n_1$ gauge fields $A_{\mu}$, $n_{1/2}$ Weyl or Majorana spinor fields $\psi$, $n_0$ two-derivative (KG) scalars $\chi$, and $n_0'$ four-derivative (FT) scalars $\varphi$, all conformally coupled to a classical curved spacetime background with metric $g_{\mu\nu}$ (and tetrad $e_{\mu}^{\,a}$):

\begin{eqnarray}
  \label{S_mat}
    S_{{\rm mat}}&=&\int d^4x\sqrt{g} \biggr[\frac{1}{4}F_{\mu\nu}F^{\mu\nu}+i\bar{\psi}e^{\mu}_{\,a}\gamma^a D_{\mu}\psi
    \nonumber\\
    &+&\frac{1}{2}\chi(-\Box+\frac{1}{6}R)\chi+\frac{1}{2}
    \varphi\Delta_4\varphi\biggr]
\end{eqnarray}
with $\Delta_4\!=\! \Box^2+(2R^{\mu\nu} - \frac{2}{3}R g^{\mu\nu})\nabla_{\mu} \nabla_{\nu} + \frac{1}{3}(\nabla^{\mu}R)\nabla_{\mu}$ (the Fradkin-Tseytlin-Paneitz operator \cite{Fradkin:1982xc, Fradkin:1981jc, paneitz2008quartic}).   
This action is invariant under the Weyl symmetry $g_{\mu\nu}\to\Omega^{2}g_{\mu\nu}$, $e_{\mu}^{\,a}\to\Omega^{1} e_{\mu}^{\,a}$, $A_{\mu}\to\Omega^{0}A_{\mu}$, $\psi\to\Omega^{-3/2}\psi$, $\chi\to\Omega^{-1}\chi$, $\varphi\to\Omega^{0}\varphi$.

To $S_{{\rm mat}}$, we add the gravitational action
\begin{eqnarray}
\label{S_grav}
    S_{{\rm grav}}&=&\int d^4x\sqrt{g} \biggr[\frac{1}{16\pi  G}(R+2\Lambda)  \nonumber \\ && + \lambda_1 C^2 + \lambda_2 E + \lambda_3 R^2 + \lambda_4\Box R\biggr]\quad
\end{eqnarray}
where $R$ is the Ricci scalar, $\Lambda$ is the cosmological constant, $C^2=R_{\alpha\beta\gamma\delta}^{2}-2R_{\alpha\beta}^{2}+\frac{1}{3}R^2$ is the square of the Weyl tensor, and $E=R_{\alpha\beta\gamma\delta}^{2}-4R_{\alpha\beta}^{2}+R^{2}$ is the Euler density (or Gauss-Bonnet term).\footnote{The couplings $\lambda_1, \lambda_2, \lambda_3$ are sometimes denoted $\lambda_1 = 1/\lambda$, $\lambda_2 = \theta/\lambda$, and $\lambda_3 = -\omega/3\lambda$ in the literature \cite{Codello:2006in, deBerredo-Peixoto:2004cjk}.}  This local action, including only terms up to four derivatives, suffices to renormalize the UV divergences arising from quantizing the matter and to determine all of the corresponding gravitational $\beta$-functions.

\section{Conformal Wick rotation}
\label{conformal_wick}

To renormalize the theory, we analytically continue to Euclidean spacetime in which the metric has a definite signature. Conventionally, QFT is studied in maximally symmetric spacetimes (Minkowski, dS or AdS) where the continuation is obvious. However, the real universe is instead approximated by an FRW line element $ds^2 = a^{2}(\tau)(-d\tau^{2}+d\Omega_{K}^{2})$ where only the {\it spatial} line element $d\Omega_{K}^{2}$ is maximally symmetric. As shown in \cite{Turok:2022fgq, Boyle:2022lcq}, the correct continuation depends on the values of the conserved cosmological parameters.  For {\it realistic} values, the universe expands from a radiation-dominated Big Bang in the past to an asymptotically de-Sitter boundary in the future. The appropriate Wick rotation is then {\it conformal}: rotating the conformal time to imaginary values $\tau\to-i\tau_{E}$ also rotates the conformal scale factor $a\to-ia_{E}$ \cite{Turok:2022fgq}.  (This is most easily seen near the bang, where the universe is radiation-dominated and $a(\tau)\propto\tau$.) The difference between an ordinary Wick rotation in flat spacetime and a conformal one in cosmology is dramatic: the former yields a coefficient $-1/16\pi G$ in the Euclidean Einstein-Hilbert action whereas the latter yields $+1/16\pi G$. 
This is why, for positive $G$, the Einstein-Hilbert term in Eq.~(\ref{S_grav}) has a positive coefficient.~\footnote{Note that either Wick rotation (regular or conformal) yields an Euclidean Einstein-Hilbert action that is unbounded below ~\cite{Gibbons:1978ac}.  This poses a well-known difficulty if one wants to path integrate over $g_{\mu\nu}$; but in this paper we treat gravity as a classical background and only path integrate over the matter fields.}

\section{Gravitational Beta Functions}  
\label{beta_fn_sec}

The exact renormalization group equation (ERGE~\cite{wetterich1993exact}) provides an elegant framework for calculating gravitational beta functions~\cite{Percacci:2017fkn, Percacci:2005wu}. 
Instead of the usual effective action $\Gamma$, obtained by integrating over all field modes, one considers $\Gamma_{k}$, obtained by only integrating over the high momentum modes $q\gtrsim k$. It obeys \cite{wetterich1993exact}
\begin{equation}
  \label{ERGE}
  \partial_{t}\Gamma_{k}=\frac{1}{2}
  {\rm Tr}\left[\left(\frac{\delta^{2}\Gamma_{k}}{\delta\Phi\delta\Phi}+R_{k}\right)^{-1}\partial_{t}R_{k}\right]
\end{equation}
where $\Phi$ stands for all fields over which we are path integrating, $t={\rm ln}\,k$ and the trace is the formal sum over modes. Each inverse propagator $z$ is replaced by a modified propagator $P_{k}(z)=z+R_{k}$ that suppresses the propagation of modes with $q \lesssim k$. 
The advantage of dealing with the flow of the action with $k$ rather than the action itself, is that the flow is finite. Furthermore, should the flow reveal a UV fixed point, it follows that the high frequency modes cancel out so that the full effective action $\Gamma$ is actually UV finite.  

To evaluate the rhs of (\ref{ERGE}) for the theory defined by $S_{mat}+S_{grav}$, the KG scalar's inverse propagator $z_{0}= -\Box+\frac{R}{6}$ is replaced by $P_k^{(0)}(z_0) = z_0  + R_k(z_0)$, with $R_k(z)=(k^2-z)\Theta (k^2-z)$ the ``optimized cut-off"~\cite{Litim:2001up}.  Similarly, for Dirac spinors, $z_{1/2} = -\Box + \frac{R}{4}$ (the square of the Dirac operator), for gauge fields $z_{1} = -\Box\delta^{\mu}_{\nu} + R^{\mu}_{\nu}$, and for ghosts $z_{gh}= -\Box$ (see {\it e.g.}, \cite{Percacci:2005wu}). Adding the optimized cutoff, we obtain $P_{k}^{(1/2)}$, $P_{k}^{(1)}$ and $P_{k}^{(gh)}$, respectively. For FT scalars, $z_{0'}=\Delta_4$, for which (by dimensions) the optimized cutoff is $R_k(z)=(k^4-z)\Theta (k^4-z)$ and, as before, $P_{k}^{(0')}$ is their sum.
 
In this notation, the ERGE (\ref{ERGE}) becomes
\begin{eqnarray}
  \label{ERGE with matter}
    &\partial_t \Gamma_k\!=\!\frac{n_{0}}{2}\text{Tr}\frac{\partial_t P_{k}^{(0)}(z)}{P_{k}^{(0)}(z)}-\frac{n_{1/2}^{D}}{2}\text{Tr}\frac{\partial_t P_{k}^{(1/2)}(z)}{P_{k}^{(1/2)}(z)}\qquad\qquad& \\
    &\!+\!\frac{n_{1}}{2}\text{Tr}\frac{\partial_t P_{k}^{(1)}(z)}{P_{k}^{(1)}(z)}\!-\! n_{1}\text{Tr}\frac{\partial_t P_{k}^{(gh)}(z)}{P_{k}^{(gh)}(z)}\!+\!\frac{n'_0}{2}\text{Tr}\frac{\partial_t P_{k}^{(0')}\!(z)}{P_{k}^{(0')}\!(z)}&\nonumber
\end{eqnarray}
for $n_0$ KG scalars, $n^{D}_{1/2}=\frac{1}{2}n_{1/2}$ Dirac spinors (half the number $n_{1/2}$ of Weyl or Majorana spinors), $n_1$ gauge bosons, and $n_0'$ FT scalars. The spinor and ghost terms have minus signs due to fermionic statistics and there are two ghosts per gauge boson to cancel the two unphysical polarizations. 

To evaluate the traces in Eq.~(\ref{ERGE with matter}), we use heat kernel methods. If $A$ is a positive elliptic differential operator on a $d$-dimensional Riemannian manifold, the trace of a function $f(A)$ is given by
\begin{equation}
  \label{Tr_f_v1}
  {\rm Tr}\,f(A)=\sum_{i}f(\lambda_i)
  =\int_{0}^{\infty}dt\,K(A,t)\,
  \tilde{f}(t)
\end{equation}
where $\lambda_i$ are $A$'s eigenvalues, $K(A,t)\equiv \sum_{i}{\rm e}^{-t\lambda_{i}}$ is the trace of the heat kernel of $A$, and $f(\lambda)=\int_{0}^{\infty}dt {\rm e}^{-\lambda t}\tilde{f}(t)$, with $\tilde{f}$ its Laplace transform.  The heat kernel expansion for a $p$th-order operator $A$ is \cite{Vassilevich:2003xt, Gusynin:1989ky} 
\begin{equation}
  \label{K}
  K(A,t)=\sum_{m\geq0}B_{2m}(A)t^{-n}\quad
  \left(\!n\equiv\frac{d-2m}{p}\right)
\end{equation}
with $B_{2m}(A)\!=\!\int d^{d}x\sqrt{g}\,b_{2m}(A)$ the Seeley-DeWitt coefficients~\cite{Vassilevich:2003xt, Gusynin:1989ky}. So the trace (\ref{Tr_f_v1}) becomes
\begin{equation}
  \label{Tr_f_v2}
  {\rm Tr}\,f(A)=\sum_{m\geq0}
  B_{2m}(A)Q_{n}^{}(f)
\end{equation}
where $Q_{n}(f)\equiv\int_{0}^{\infty}dt\,t^{-n}\tilde{f}(t)$. One can check that
\begin{equation}
      Q_{n}(f)=\left\{\begin{array}{ll}
      \frac{1}{\Gamma(n)}\int_{0}^{\infty} dz\,z^{n-1}f(z) & (n > 0) \\
      (-1)^{n}f^{(|n|)}(0) & (n\leq 0)
      \end{array}\right.
\end{equation}
where $f^{(|n|)}(z)$ is the $|n|$th derivative of $f(z)$. 

For $f=\frac{\partial_{t}P_{k}(z)}{P_{k}(z)}$, with $P_{k}(z)=z+R_{k}(z)$ and $R_{k}(z)=(k^{p}-z)\Theta(k^{p}-z)$ (recall, $p=2$ for all fields except FT scalars, for which $p=4$), we have $\partial_{t}P_{k}(z)=pk^{p}\Theta(k^p-z)$. It follows that
\begin{equation}
  Q_{n}(\frac{\partial_{t}P_{k}}{P_{k}})=\left\{\begin{array}{ll}
  p\,k^{np}/\Gamma(n+1)&\quad(n\geq 0) \\
  0 &\quad(n < 0) \end{array}\right.
\end{equation}
so that (with $d=4$), Eq.~(\ref{Tr_f_v2}) becomes
\begin{equation}
  {\rm Tr}\frac{\partial_{t}P_{k}(z)}{P_{k}(z)}
  =\frac{pk^{4}\!B_{0}(z)}{\Gamma\Big(\frac{4}{p}\!+\!1\Big)}
  \!+\!\frac{pk^{2}\!B_{2}(z)}{\Gamma\Big(\frac{2}{p}\!+\!1\Big)}
  \!+\!pB_{4}(z).
\end{equation}
This result gives all terms on the rhs of (\ref{ERGE with matter}).  The Seeley-Dewitt coefficients $B_{i}(z)$ for the 2nd-order operators $z_0$, $z_{1/2}$, $z_{1}$ and $z_{gh}$ may be obtained from \cite{Christensen:1978md, Vassilevich:2003xt}\footnote{In particular, see Eqs.~(4.26-4.28) in  \cite{Vassilevich:2003xt}, which are useful for checking the results in Table 3 of \cite{Christensen:1978md} and Table 1 of \cite{Vassilevich:2003xt}, particularly in the spin 1/2 and spin 1 cases.} and the $B_{i}(z)$ for the 4th-order operator $z_{0}'$ from Eqs.~(28,29) in \cite{Gusynin:1989ky}.

Contributions to the lhs of (\ref{ERGE with matter}) arise from the running of the gravitational effective action:
\begin{eqnarray}
 \partial_{t}\Gamma_{k}&=  \int d^4x\sqrt{g} \biggr[\frac{1}{16\pi}(\beta_{1/G}^{}R+2\beta_{\Lambda/G})\qquad& \nonumber\\
    &\quad+ \beta_1C^2 + \beta_2E+\beta_3R^2+\beta_4\Box R\biggr],\qquad&
\end{eqnarray}
with each beta function the $t$-derivative of the corresponding coupling in (\ref{S_grav}). Thus, Eq.~(\ref{ERGE with matter}) yields
\begin{eqnarray}
\label{rges}
        \beta_{\Lambda/G} &=& \frac{k^4}{4\pi}\biggr(n_0-2n_{1/2}+2n_1+2n'_{0}\biggr)\nonumber\\
        \beta_{1/G} &=& \frac{k^2}{6\pi}\biggr(n_{1/2}-4n_1+n'_0\biggr)\nonumber\\
        \beta_1 &=& \frac{\frac{3}{2}n_0+\frac{9}{2}n_{1/2}+18n_1-12n'_0}{(4\pi)^2 180}\nonumber\\ 
        \beta_2 &=& \frac{-\frac{1}{2}n_{0}-\frac{11}{4}n_{1/2}-31n_1+14n'_0}{(4\pi)^2 180}\nonumber\\
        \beta_3 &=& 0 \nonumber\\ 
        \beta_4 &=& \frac{n_0+3n_{1/2}-18n_1+12n'_0}{(4\pi)^2 180}\,,
\end{eqnarray} 
whose solutions are straightforwardly obtained:
\begin{eqnarray}
  \label{Lambda_G_running}
  \frac{\Lambda(k)}{G(k)}&=&
  \frac{\Lambda(0)}{G(0)}+(n_{0}-2n_{1/2}+2n_{1}+2n_{0}')
  \frac{k^{4}}{16\pi}\,,\nonumber\\
  \frac{1}{G(k)}&=&\frac{1}{G(0)}+
  (n_{1/2}-4n_{1}+n_{0}')\frac{k^{2}}{12\pi}\,,\nonumber\\
    \lambda_i(k) &=& \lambda_i(k_0) + \beta_i ln(k/k_0), \quad i=1,\dots, 4
    \end{eqnarray} 
with $k_0$ an arbitrary scale. 

\section{Standard Model Implications}

We have found the running of all terms in the gravitational action by integrating out conformally coupled free fields: $n_0$ KG and $n_{0'}$ FT scalars, $n_{1/2}$ Weyl or Majorana fermions and $n_1$ gauge bosons. 

What does this imply for the SM? The SM is, of course, an interacting QFT. However, when extrapolated to ultra-high energies (up to the Planck scale and well beyond) SM matter-matter couplings are small and perturbative~\cite{Buttazzo:2013uya}, so a free field approximation is not entirely unreasonable.  

The SM contains $n_1=8+3+1=12$ gauge bosons, $n_0=4$ real KG scalars, $n_{0'}=0$ FT scalars, and either $n_{1/2}=3\times15=45$ or $n_{1/2}=3\times16=48$ Weyl fermions (depending on whether or not we include right-handed neutrinos\footnote{RH neutrinos provide the simplest renormalizable explanation for the observed neutrino masses and oscillations \cite{ParticleDataGroup:2024cfk}; and they explain the dark matter \cite{Canetti:2012kh, Boyle:2018tzc, Boyle:2018rgh, Shaposhnikov:2024nhc} and  cosmological matter-antimatter asymmetry \cite{Buchmuller:2004nz, Davidson:2008bu}, which are both unexplained in the RH-neutrinoless SM. \label{RHnu}}).  
The first implication of (\ref{Lambda_G_running}) is that the vacuum energy (or cosmological constant) term diverges to minus infinity in the UV. Heuristically, this would give spacetime a huge negative ``tension" (energy per 3-volume), causing it to be highly negatively curved on short distances. However, the Lorentzian continuation of a ``cosmological constant" term $\propto k^4$, with $k$ the {\it Euclidean} cutoff, is far from clear. The resulting divergences violate Lorentz invariance~\cite{Koksma:2011cq} and the scale dependence suggests the effective cosmological constant varies strongly with cosmological epoch, in conflict with observation.  (For an attempt to understand these issues in de Sitter spacetime, see Ref.~\cite{Ferrero:2025ugd}.)  The most straightforward interpretation, which we explore here, is that such QFT divergences simply must cancel in order that the matter couples consistently to gravity. 

Second, since (if we include right-handed neutrinos) $n_{1/2}=4 n_1$  and $n_{0'}=0$ in the SM, Newton's constant $G$ does {\it not} run in the UV. This is bad news because the effective dimensionless coupling in graviton exchange $\tilde{G}(k)=k^{2}G(k)$ diverges so that perturbative unitarity is violated~\cite{Han:2004wt,Caron-Huot:2024lbf}.\footnote{If we don't include RH neutrinos, things are even worse: in addition to the issue in footnote \ref{RHnu}, $G$ develops a pole at the Planck scale, and becomes negative beyond it, again indicating the theory is sick at high energies.} Finally, since $\lambda_1$ and $\lambda_2$ are not fixed (and moreover diverge), the matter stress-tensor correlators $\langle T\ldots T\rangle$ as inferred from the effective gravitational action, have UV divergences, as we detail below. For all these reasons, the minimal SM does not seem to couple consistently to gravity in the UV. 

\section{The Fixed Point Theories}
\label{Sec:Running}

In the literature on asymptotic safety, {\it e.g.}~\cite{weinberg1979ultraviolet, Percacci:2017fkn}, it is conventional to convert any dimensionful couplings (in our case $\Lambda$ and $G$) to dimensionless parameters ($\tilde{\Lambda}=k^{-2}\Lambda$ and $\tilde{G}= k^2 G$), and to study their running and fixed points. From (\ref{rges}), we see that $\tilde{\Lambda}$ and $\tilde{G}$ {\it do} possess UV fixed points,
\begin{eqnarray}
    \label{Lambda_star}
    \tilde{\Lambda}_{*}&=&
    \frac{3(n_{0}-2n_{1/2}+2n_{1}+2n_{0}')}{4(n_{1/2}-4n_{1}+n_{0}')} \nonumber\\
    \label{G_star}
    \tilde{G}_{*}&=&\frac{12\pi}{(n_{1/2}-4n_1+n'_0)}\,.
\end{eqnarray}
Next we can define the rescaled RG parameter:
\begin{equation}
    \tilde{k}^2 = \frac{k^2}{\tilde{G}_*} = \frac{1}{G(k)},
\end{equation}
which should be interpreted as the cut-off measured in units of running Planck mass \cite{Coleman:1985rnk, Codello:2012sn}. The gravitational action (\ref{S_grav}) then reads:
\begin{eqnarray}
\label{S_gravUV}
    S_{{\rm grav}}&=&\int d^4x\sqrt{g} \biggr[\frac{\tilde{k}^2}{16\pi}(R+2\tilde{k}^2\lambda)  \nonumber \\ && + \lambda_1 C^2 + \lambda_2 E + \lambda_3 R^2 + \lambda_4\Box R\biggr]
\end{eqnarray}
where we have defined the dimensionless coupling $\lambda = \tilde{\Lambda}\tilde{G} = \Lambda G$.
Written in this way, the action only depends on the four dimensionless couplings $\lambda, \lambda_1, \lambda_2, \lambda_3$ (we ignore $\lambda_{4}$ since $\Box R$ is an irrelevant total derivative term), 
and the explicit integer powers of $\tilde{k}$ dictated by dimensional analysis (with no other dimensionful couplings). 

Now we observe that the four non-trivial couplings in this theory ($\lambda$, $\lambda_1$, $\lambda_2$, $\lambda_3$) have a simultaneous fixed point, and the fixed value of $\lambda$ is set to zero, provided that the matter fields are conformally coupled and the field content satisfies
\begin{equation}
  \label{matter_content}
    n_{1/2}=4n_{1},\;\;n_0'=3n_1,\;\;n_0=0.
\end{equation}
Note four key points about this condition:

1. Eq.~(\ref{matter_content}) (needed to make the beta functions $\beta_{\Lambda/G}$, $\beta_{1}$, $\beta_{2}$, and $\beta_{3}$ all vanish) also precisely implies that all stress tensor correlators $\langle T\ldots T\rangle$ are free of UV divergences. This follows from the fact that the stress tensor is given by the metric variation of the matter action, {\it i.e.}, the action {\it excluding} the Einstein-Hilbert term; and, similarly, the stress tensor {\it correlators} $\langle T\ldots T\rangle$ are obtained by varying the effective action $\Gamma$ (again excluding the Einstein-Hilbert term).   
In particular, requiring no UV divergence in: $\langle T_{\mu\nu}\rangle$ requires $\beta_{\Lambda/G}=0$; $\langle T^{\alpha}_{\alpha}(x)T_{\mu\nu}(y)\rangle$ requires $\beta_{3}=0$ (Eq. 8.4 in \cite{Osborn:1993cr}), $\langle T_{\mu\nu}(x)T_{\rho\sigma}(y)$ also requires $\beta_{1}=0$ (Eqs. 8.8, 8.12 in \cite{Osborn:1993cr}); and $\langle T_{\mu\nu}(x)T_{\rho\sigma}(y)T_{\kappa\lambda}(z)\rangle$ also requires $\beta_{2}=0$ (8.26, 8.34 in \cite{Osborn:1993cr}). As these restrictions remove all divergences in the matter effective action, {\it all} correlators $\langle T\ldots T\rangle$ are free of UV divergences.  (Note that the divergences are {\it universal} - if they cancel in the vacuum, they cancel in any state.)

\begin{figure}
    \centering
    \includegraphics[width=1\linewidth]{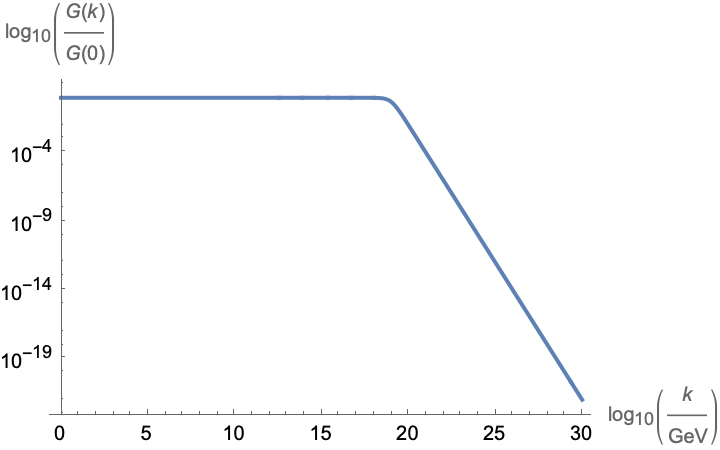}
    \caption{Running of Newton's gravitational constant $G$ with cut-off scale $k$, for free fields satisfying (\ref{matter_content}).}
    \label{Newton's running}
\end{figure}

2. This set of free fields leaves us with a) a finite cosmological constant and Newton's constant of arbitrary magnitude (set by observations) in the IR, and b) a scale-invariant theory (\ref{S_gravUV}) at the UV fixed point with effective gravitational coupling (Newton's constant $G$) ``softening" as $k^{-2}$ with a break at $k\sim m_{Pl}$ (shown in Fig.~\ref{Newton's running}).  

3. The coefficient of $R$, {\it i.e.}, the inverse of Newton's constant, diverges in the UV.  However, in situations where the matter is dominated by a conformal radiation (with a traceless stress tensor) -- {\it e.g.}\ at the Big Bang -- $R$ vanishes by the equations of motion, hence the action remains finite.

4. Finally, Eq.~(\ref{matter_content}) is striking since, in the standard model ($n_{1}=12$), it requires $n_{1/2}=48$, which is automatically satisfied by three generations of standard model fermions (including right-handed neutrinos)!  The price of all these cancellations is twofold: we {\it must} include $3 n_1=36$ FT scalars, and we {\it must not} include {\it any} fundamental KG scalars.  We discuss these two points in the next section.

\section{Discussion}

In this paper, we have studied free, conformally-coupled quantum matter fields. We  have identified a special class of such theories (\ref{matter_content}) exhibiting gravitational UV fixed points with intriguing and encouraging properties. 
These fixed points are certainly interesting from a theoretical standpoint, but fundamental questions remain about whether/how they might relate to the real world.  Here we point out two key questions, and mention some related speculations.

Eq. (\ref{matter_content}) requires $n_{0}=0$ KG scalars, and so appears to suggest that the SM Higgs (a KG scalar) is not a fundamental field.  One possibility is that it is instead a composite field formed from FT scalars and other SM fields.  Since interacting FT scalars are asymptotically free \cite{Holdom:2023usn, Holdom:2024cfq}, one appealing possibility is that the weak scale emerges quantum mechanically in the same way that the QCD scale does, as the scale where an asymptotically free coupling becomes strong.  If this picture can be realized, it offers an appealing solution to the gauge hierarchy problem~\cite{BT} (for related ideas, see \cite{Romatschke:2024hpb}).

For the standard model (with its $n_{1}=12$ gauge bosons), the fixed point (\ref{matter_content}) also requires $n_{0}'=36$ FT scalars.  
The Euclidean action for FT scalars, even interacting, asymptotically free FT scalars of the kind mentioned above, is positive definite. So there is no problem with including FT scalars in the Euclidean path integral as we have done here.  In fact, Costello \cite{Costello:2021bah} and Bittleston et al \cite{Bittleston:2022nfr} have argued that such FT scalars {\it must} be included, for anomaly-cancellation reasons, to make sense of certain interesting theories in 4D spacetime that are dual to local holomorphic field theories on twistor space~\cite{Costello:2021bah, Bittleston:2022nfr}. The question of how to analytically continue such FT scalar theories to {\it Lorentzian} signature is a topic of lively debate (see {\it e.g.}~\cite{Lee:1969fy, Lee:1970iw, Bogolubov:1990ask, Hawking:2001yt, Rivelles:2003jd, Bender:2007wu, Salvio:2015gsi, Donoghue:2017fvm, Donoghue:2019fcb, Donoghue:2019ecz, Donoghue:2021eto, Donoghue:2021meq, Boyle:2021jaz, Tseytlin:2022flu, Lehners:2023fud, Turok:2023amx, Holdom:2023usn, Holdom:2024cfq}). These arguments will be reviewed and addressed in a forthcoming publication~\cite{BT}.

\section{Acknowledgements}
We thank Sam Bateman, Roland Bittleston, John Donoghue, Astrid Eichorn, Roberto Percacci, Alessia Platania and Roman Zwicky for helpful discussion. VV is supported by the School of Physics and Astronomy Studentship at the University of Edinburgh. LB and NT are supported by STFC Consolidated Grant ‘Particle Physics at the Higgs Centre,’ and  NT is supported by the Higgs Chair at the University of Edinburgh. Perimeter Institute is supported by the Government of Canada, via Innovation, Science and Economic Development, Canada and by the Province of Ontario via the Ministry of Research, Innovation and Science.

\bibliography{references}

\end{document}